# Mechanisms and kinetics of C-S-H nucleation approaching the spinodal line: Insights into the role of organics additives

Christophe Labbez[1,*], Lina Bouzouaid[1], Alexander E.S. Van Driessche[2], Wai Li Ling[3], Juan Carlos Martinez[4], Barbara Lothenbach[5], Alejandro Fernandez-Martinez[2,*]

[1] *ICB, UMR 6303 CNRS, Univ. Bourgogne Franche-Comté, FR-21000 Dijon, France*
[2] *Univ. Grenoble Alpes, Univ. Savoie Mont Blanc, CNRS, IRD, IFSTTAR, ISTerre, FR-38000 Grenoble, France*
[3] *Univ. Grenoble Alpes, CEA, CNRS, IRIG, IBS, FR-38000 Grenoble, France.*
[4] *Non-Crystalline Diffraction Beamline−Experiments Division, ALBA Synchrotron Light Source, Cerdanyola del Vallès, SP-08290 Barcelona, Spain*
[5] *Empa, Concrete & Asphalt Laboratory, CH-8600 Dubendorf, Switzerland*

* Christophe.Labbez@u-bourgogne.fr, Alex.Fernandez-Martinez@univ-grenoble-alpes.fr

## Abstract

Wet chemistry C-S-H precipitation experiments were performed under controlled conditions of solution supersaturation in the presence and absence of gluconate and three hexitol molecules. Characterization of the precipitates with SAXS and cryo-TEM experiments confirmed the presence of a multi-step nucleation pathway. Induction times for the formation of the amorphous C-S-H spheroids were determined from light transmittance. Analysis of those data with the classical nucleation theory revealed a significant increase of the kinetic prefactor in the same order as the complexation constants of calcium and silicate with each of the organics. Finally, two distinct precipitation regimes of the C-S-H amorphous precursor were identified: i) a *nucleation regime* at low saturation indexes (*SI*) and ii) a *spinodal nucleation regime* at high *SI* where the free energy barrier to the phase transition is found to be of the order of the kinetic energy or less.

# 1. Introduction

In Portland cement, calcium silicate hydrate (C-S-H) is the principal hydration product of alite ($C_3S$) and belite ($C_2S$), playing a decisive role in the final properties of concrete [1]. It is generally accepted that the formation of this hydrate follows a dissolution-precipitation process [2,3]. To control the workability and final properties of concrete, a large variety of additives are added to the cement mixture, such as setting retarders, accelerators and superplasticizers, among others [4-7]. These chemicals alter different processes occurring in cementitious systems, including their hydration behavior. Extensive experimental work has shown that numerous additives have a retardation effect on the hydration of $C_3S$ [8-13], which has often been interpreted as a consequence of the disruption of the nucleation and/or growth process of C-S-H. This, however, has never been demonstrated. So far it is known that some retarders, such as hexitols, yellow dextrin and cellulose ethers [12-17], do not limit the dissolution of $C_3S$, while others, like polycarboxylate-type superplasticizer [18] and gluconate [19], do. Thus, in order to develop more efficient cement additives, a detailed understanding of the initial formation of C-S-H in their absence and presence is needed.

In a recent work by Krautwurst et al. [20], time-resolved potentiometry coupled with transmittance measurements, small angle X-ray scattering and cryo-TEM experiments were used to study the homogeneous nucleation of C-S-H from diluted solutions. It was found that C-S-H formation follows a two-step pathway. In the first step, a decrease in the solution's transmittance signals the formation of amorphous spheroids. During a second step, these spheroids then aggregate and crystallize into typical sheet-shaped C-S-H particles, indicated by another decrease in transmittance [20]. In another study, Plank et al. [54] used TEM imaging to show the presence of metastable C-S-H nanoparticles with globular morphology and sizes between 20 and 60 nm, similar to those observed by Krautwurst et al. [20], during the initial stages of the nucleation reaction. After less than one hour, those nanoparticles were observed to grow gradually into the characteristic nanofoils of early C-S-H.

These two studies described the formation of C-S-H through a so-called 'non-classical' nucleation pathway, which has been observed for a large variety of organic and inorganic systems, but remains a much-debated topic due to the complex processes involved [21,22]. The formation of precursor metastable phases is predicted by Ostwald 's rule of stages, invoking the low energetic cost from a kinetic point of view to form the less stable phase (i.e. more soluble) from a supersaturated solution. Many recent observations have contributed lately to a more complete description of this rule of stages: several authors have described the formation of amorphous or nanoparticle precursors (e.g. [23-25]), which in some cases act as the building blocks of the crystalline phases through complex diffusion and aggregation processes which are still poorly understood (e.g. [26]). These observations have been validated for several systems ($CaCO_3$, $CaSO_4$, C-S-H, etc.), but are still far from being fully understood, and a universal law describing the full crystallization process is still far from being reached (e.g. [27,28]). Moreover, Plank et al. showed that polycarboxylate ether-based superplasticizers (PCEs) control the kinetics of the transition from globular to nanofoil-like C-S-H [54]. Indeed, in the presence of those additives, this conversion is strongly delayed. It is argued that PCEs form a layer around C-S-H globules in a core-shell geometry. The authors hypothesized that this external layer slows down the water diffusion into the particles' core, and consequently decreases the kinetics of dissolution and re-crystallization of the core particles. Yet, most of the commonly employed additives are small simple organics, such as gluconate. This compound is used routinely by the concrete industry as retarder to control the setting time of cement. [30-34]. Gluconate is also efficient in increasing the compressive strength of concrete [33,34]. Another saccharide derivative such as sorbitol (a neutral sugar alcohol) is also used as a set retarder for Portland cement: when added in the material at 0.40 wt%, sorbitol delays the setting for 2 days [35]. Sorbitol also serves as a water-reducing plasticizer, decreasing final porosity and giving greater mechanical strength and durability [36].

Overall, C-S-H nucleation from diluted solutions is an archetypal case of multi-step nucleation, in which a metastable phase (amorphous spheroids) persists for a long time (~minutes to hours) before the final crystalline phase emerges. The aggregation of these spheroids has been suggested as a critical step [20], concomitant to the crystallization, but further investigations are needed to fully characterize the precipitation process, in particular in the presence of additives.

Therefore, the focus of this work is on the nucleation of the amorphous C-S-H spheroids in the presence of different cement additives, D-sorbitol, D-mannitol, D-galactitol and gluconate, that act as retardants of $C_3S$ hydration. The first three molecules share the same chemical formula but have different stereochemistry. Precipitation experiments were performed in diluted aqueous solutions and the pH of the experiments was set at the value close to the one observed in a typical cement paste (pH 12.8). As opposed to the conditions reigning in real cement, and with the objective of simplifying the study, our goal was to study homogeneous nucleation, so the solutions were kept free from grain/impurities, in order to avoid heterogeneous nucleation. First, the effects of additives on the nucleation kinetics of C-S-H were studied by determining the induction time as a function of supersaturation using light transmittance measurements. Secondly, small angle X-ray scattering (SAXS) experiments were performed in order to characterize the size and shape distribution of the C-S-H nanoparticles formed during precipitation, in the absence and presence of additives. Finally, cryo-TEM, was used to get a clearer picture of the particle morphology in the presence and absence of organic molecules.

## 2. Material and methods

*2.1. Materials*

The different stock solutions were prepared by dissolving $CaCl_2$ $2H_2O$ (Sigma-Aldrich, ≥99% purity), sodium gluconate ($C_6H_{11}NaO_7$, Sigma-Aldrich, ≥99% purity) D-sorbitol ($C_6H_{14}O_6$, Sigma-Aldrich, ≥99% purity), D-mannitol ($C_6H_{14}O_6$, Sigma-Aldrich, ≥99% purity), and D-galactitol ($C_6H_{14}O_6$, Sigma-Aldrich, ≥99% purity) and sodium chloride (NaCl, Sigma-Aldrich, ≥99% purity), in boiled and degassed milliQ water.

The hexitols used in this study are all isomers, sharing the formula, $HOCH_2(CHOH)_2CH_2OH$, but differ in the stereochemical arrangement of the OH groups as illustrated in Figure 1.

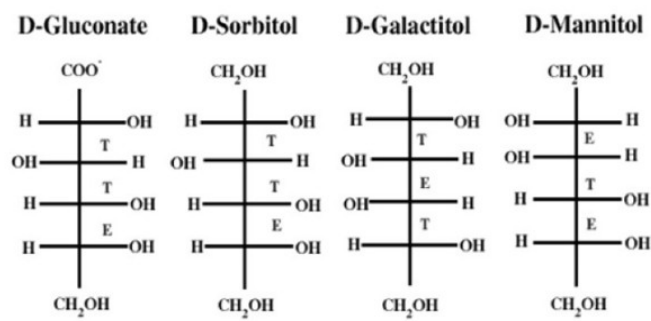

**Figure 1.** Stereochemistry of the different organic molecules used in this study.

*2.2. Homogeneous precipitation of C-S-H*

The precipitation of $CaO\text{-}SiO_2\text{-}H_2O$ was followed using an automated titrator instrument (Metrohm 905 Titrando) by mixing $CaCl_2$ and $Na_2SiO_3$ solutions. All measurements were performed in a thermostated reactor at 23.0 ± 0.1°C. The solution in the reactor was continuously stirred with a magnetic stirrer at a constant rate of 430 rpm. A nitrogen flow circulated continuously above the solution to avoid contamination with $CO_2$. It was taken care that the gas did not enter the solution to avoid any disturbance of the electrodes. In the reactor, 150 mL of a solution containing 25mM NaCl background electrolyte was placed together with a fixed concentration of the organic of interest, between 0.1mM and 10mM, 80 mM of NaOH and 15 mM of $CaCl_2.(H_2O)_2$. After 3 minutes of equilibration time, 3 mL of a titrant solution containing a silicate concentration was added at a rate of 1.2 mL/min. In the reactor, the initial silicate concentrations were varied between 0.05 mM and 0.25 mM and the solution pH between 12.7 and 12.8.

In order to monitor the formation of the first particles of C-S-H in absence and in presence of the small organics, an optical electrode (Metrohm Optrode, 6.1115.000) measuring the light transmittance of the system at a wavelength of 610 nm was used. The later was used to determine the induction time of nucleation ($t_{ind}$), that is the time needed to observe the first nuclei. It should be stressed that $t_{ind}$ may vary with the experimental detection technique used, see below. The pH was determined with a pH electrode (Metrohm pH Unitrode with Pt 1000, 6.0259.100), which allows reliable measurements up to pH = 14. The pH electrode was calibrated prior to the measurements with standard buffer solutions (pH 10, 12.45 and 14 from Sigma Aldrich).

*2.3. Saturation index calculation.*

The saturation index, *SI*, was calculated with a speciation model solved by the geochemical software PHREEQC version 3 (3.6.2-15100) using the WATEQ4f database completed with the Ca-organics-silicate complexes for the four organics of interest that we determined in two recent studies [37,38] and the C-S-H model of Haas and Nonat [39]. The consideration of the aqueous $CaH_2SiO_4^0$ complex as suggested by some studies [58], but not in the PHREEQC database [59], had no significant effect on the derived complex formation constant and its formation was thus not considered. *SI* is calculated by comparing the activity product of the dissolved ions of C-S-H (*IAP*) with their solubility product ($K_{sp}$); $SI = \log(IAP/K_{sp})$.

*2.4. Induction time and classical nucleation theory.*

The induction time, $t_{ind}$, was measured by light transmittance and is directly related to the reverse of the nucleation rate, $J_N$, as follows:

$$t_{ind} = \frac{A}{VJ_N} \tag{1}$$

where *V* is the volume of the supersaturated solution and *A* is a constant dependent on the experimental technique used. Within the classical nucleation theory (CNT) formalism for spherical nucleus, the induction time ($t_{ind}$) can be written as a function of the solution saturation index (*SI*) as

$$\ln(t_{ind}) = \frac{16\pi v_m^2 \gamma^3}{3(k_B T)^2 (\ln 10)^2 SI^2} - \ln(C_0) \tag{2}$$

The first term on the right hand side of Eq. 2 is nothing but the Gibbs free energy barrier to the nucleation, *ΔG*. It is a function of $v_m$ the molecular volume of the formed phase, $\gamma$ the interfacial free energy, $k_B$ the Boltzmann factor, and *T* the temperature. The second term, $C_0$, is directly related to the pre-exponential kinetic factor $K_0$ of the nucleation rate, $J_N = K_0 \exp(-\Delta G/kT)$, such as $C_0 = VK_0/A$. $K_0$ is related to thermodynamic and kinetic factors such as the number of sites susceptible to act as nucleation loci, the collision rate, and the Zeldovich factor (related to the probability of successful formation of a cluster in solution), see e.g. [40]. Using Eq (1) to fit $\ln(t_{ind})$ data plotted as a function of $1/SI^2$ allows to derive the interfacial free energy of the critical nucleus and $-\ln(C_0)$, from the slope and y-intercept, respectively. For the sake of brevity, we will refer to $C_0$ as *kinetic prefactor*.

## 2.5. Cryogenic Transmission Electron Microscopy

Titration experiments, as described in 2.2, were run in the presence of gluconate and sorbitol. Aliquots (4μl) were sampled from the solution at different reaction times and deposited onto freshly glow discharged quantifoil or lacey carbon grids, and vitrified using a Thermofisher Vitrobot Mark IV system. The organics and silicate concentrations in those experiments were 7.5 mM and 0.15 mM respectively. The grids were then mounted on a Gatan 626 single-tilt cryo-transfer holder. Imaging and selected area diffraction (SAD) was performed under low-dose conditions on a Tecnai F20 microscope operating at 200 keV. Images were recorded on a Ceta CMOS camera whereas diffraction patterns were recorded on an Amsterdam Scientific Instrument CheeTah hybrid pixel camera.

## 2.6. Small Angle X-ray Scattering (SAXS)

**Table 1. Conditions used for the SAXS experiments.**

| Additive | [Si] (mM) | [Ca] (mM) | [NaCl] (mM) | [NaOH] (mM) | [additive] (mM) |
|---|---|---|---|---|---|
| Pure | 0.10 | 15 | 25 | 18 | - |
| Gluconate | 0.10 | 15 | 25 | 18 | 0.03 |
| Gluconate | 0.10 | 15 | 25 | 18 | 0.07 |
| Gluconate | 0.10 | 15 | 25 | 18 | 0.15 |
| Gluconate | 0.10 | 15 | 25 | 18 | 0.30 |

SAXS experiments were performed at the NCD beamline of the ALBA synchrotron (Barcelona, Spain). The same Metrohm system as described in section 2.2 was used to titrate a calcium chloride solution containing gluconate within a sodium orthosilicate solution (see concentrations used in Table 1). The reaction took place in the same reactor used for the transmittance experiments described in section 2.2. A peristaltic pump was used to circulate the solution in a closed loop through a 1.5 mm diameter polyamide capillary that was set up at the beam position. The residence time in the tubing and capillary is in the order of ~20 seconds. The precipitation reaction was monitored using a pH electrode and a transmittance sensor (as described in 2.2). An X-ray beam of energy E = 12.40 keV ($\lambda$ = 1.00 Å), calibrated using a silver behenate standard, was used. The scattered radiation was collected using a Pilatus 2M two-dimensional detector (981 × 1043 pixels, 172 × 172 μm pixel size) that was placed at 6.7 m from the sample to yield a q-range 0.03 < $q$ < 2.00 nm$^{-1}$. The scattered intensity was acquired every minute in acquisitions of 30 seconds to follow the kinetics of C-S-H nucleation.

## 3. Results

*3.1. C-S-H nucleation kinetics*

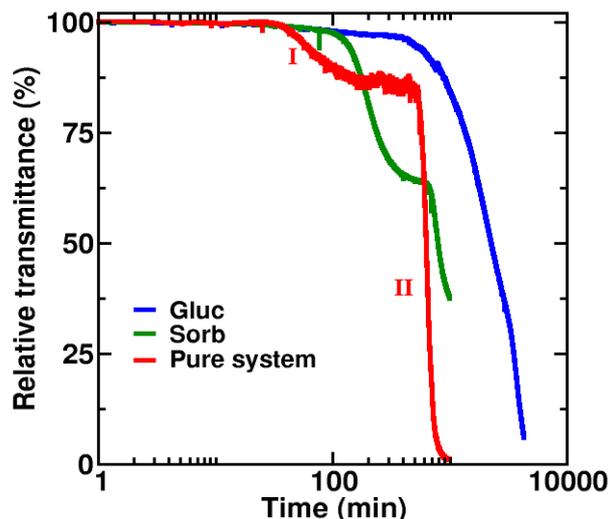

**Figure 2**. Evolution of the relative transmittance of the solution as a function of time (log-scale) for the pure C-S-H system (red line), and in the presence of 10 mM gluconate (blue) and 10 mM sorbitol (green). The initial concentrations of Si is set at 0.15 mM, Ca at 15 mM. The total solution volume is set at 153 ml, see sect. 2.2. After an initial drop of the transmittance (stage I) corresponding to the formation of amorphous spheroids, a plateau is reached. A final decrease of the transmittance is observed (stage II), corresponding to the aggregation of the amorphous spheroids and simultaneous C-S-H crystallization [20]. Note that the same behaviour is observed for the case of sorbitol but not of gluconate.

Three curves showing the typical evolution of the transmittance over time, for the pure and the 10 mM gluconate and sorbitol systems, are shown in Fig. 2. The pure system shows the same behaviour described by Krautwurst et al. [20], with two characteristic drops of the transmittance. The first drop was ascribed to the formation of amorphous spherical particles of ~50 nm size that act as precursors for the formation of nanocrystalline C-S-H; the second drop was described as resulting from a process of spheroid aggregation and changes of stoichiometry and surface chemistry, leading to massive C-S-H crystallization [20]. This behaviour is also observed in the case of sorbitol, showing two well-defined drops of the transmittance. Importantly, the case of gluconate is notably different: the signal decrease is slower than for the pure system and for the sorbitol and no clear sign of a second drop of the signal at later times is observed.

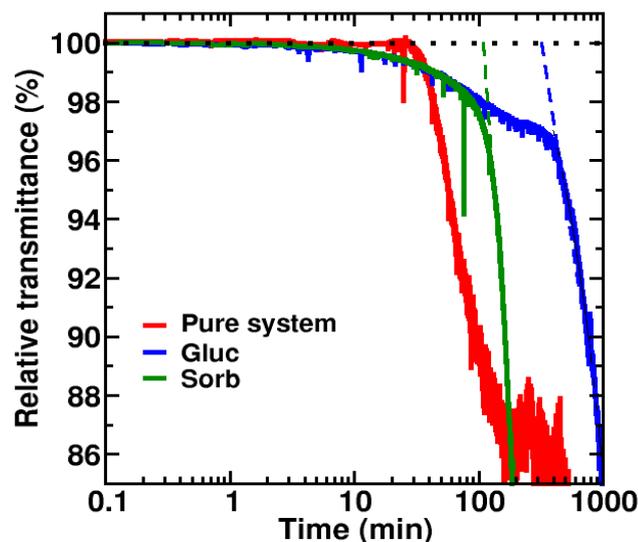

**Figure 3**. Enlargement of the initial part of the transmittance curves (Fig. 2), before the first transmittance drop of stage I, with [Si] = 0.15mM and [Ca] = 15mM. The dotted lines show the adopted fitting to determine the induction time ($t_{ind}$) as reported in Fig. 5 and defined by Eq. 2. Note the continuous decrease of the transmittance with time from the very beginning of the C-S-H precipitation experiments with gluconate and sorbitol but not in the pure system.

A detailed analysis of the initial times shows a difference between the behaviour of the pure system and that of the organic-containing systems. The neutral molecule sorbitol and the charged molecule gluconate are represented as examples in Fig. 3 in comparison with the reference system for the same initial calcium and silicate concentrations ([Si] = 0.15 mM and [Ca] = 15mM). Whereas the data from the pure system shows an initial plateau, with no changes in the transmittance until the induction time, the supersaturated solutions containing organics show an initial decrease of the transmittance that takes place immediately after mixing the reactants, followed by a stabilization step (only in the case of gluconate) prior to a second drop. This initial decrease is more pronounced in the case of gluconate than for sorbitol: in the case of sorbitol, it lasts for ~ 40 minutes, and the drop in relative transmittance is about 2%; the data from the gluconate-containing solution shows an initial decrease that lasts for ~100 minutes, and a drop in relative transmittance of ~3%.

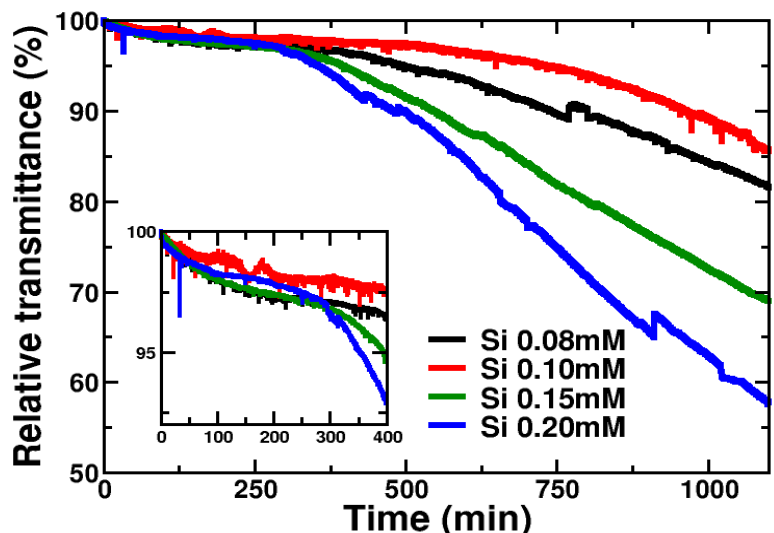

**Figure 4**. Transmittance as a function of time for the C-S-H system in presence of gluconate at 10 mM, at different supersaturation degrees: [Si] = 0.08 mM, [Si] = 0.10 mM, [Si] = 0.15 mM and [Si] = 0.20 mM.

The presence of this initial drop in the organic-containing solutions poses some questions about the interpretation of the different phenomena going on during C-S-H nucleation in the presence of additives. In the analysis by Krautwurst et al. [20] this very initial first drop in transmittance was not observed, as confirmed here (Fig. 3). The fact that this initial drop is not observed for the pure system allows making the hypothesis that it is related to the formation of aqueous Ca-Si-organic complexes. With the aim of clarifying this point, a comparison of data taken for the 10 mM gluconate system under different β-C-S-H supersaturations is presented in Figure 4 (see speciation of the supersaturated solutions in Table S1.A of the Supplementary Material). The data show that all the solutions display the same trend, with an initial drop of the transmittance in the same time interval, irrespective of the supersaturation with respect to β-C-S-H (see inset Fig. 4). Such a behaviour differs notably from that expected for a nucleation event, in which the induction time –and its associated decrease of the transmittance– responds to changes in the supersaturation of the solution. Indeed, the initial drop is independent of the supersaturation. The associated process is thus not activated. On the contrary, the second drop in transmittance (corresponding to the first drop in the pure system), observed at around ~500 min shows a dependency with the supersaturation, with faster and more pronounced decrease at higher supersaturations. This leads us to propose that the initial drop is due to the above-mentioned process of aqueous Ca-Si-organic complex formation (or formation of clusters of Ca-Si-organic complexes) and that the second drop at >500 min (Fig. 4) is a true nucleation event. Note that the first two values ([Si] = 0.08 mM and [Si] = 0.10 mM) do not follow the expected trend; this behaviour, though unexpected, can be rationalized considering the stochastic nature of the nucleation process, which leads to non-negligible uncertainties in the values of the induction times, as it will be discussed later.

The data shown in Fig. 4 allows assigning the initial drop in transmittance to a process of Ca-Si-organic complex formation. It seems that the organic molecules interfere with the nuclei precipitated, or seem to induce the formation of particular nuclei that are different in shape and size, blurring the well-defined stages observed in the reference system, in particular stage I. The presence of the complexes involving calcium, silicate, hydroxide and the organics described in more details in our previous work [38] can be an explanation in the change of the pathways observed for the pure system, and thus can be the cause of the very first decrease of the transmittance, given that it is observed at all degrees of supersaturation studied (see Fig. 4). The fact that, as shown in Fig. 3, this initial drop is larger for gluconate than for sorbitol is expected given the relative amount of calcium being part of polynuclear Ca-Si-gluconate complexes that form under these conditions, in comparison to the lower amount in hexitol complexes (see Table S1.B of the Supplementary Material).

Thus, the later drop (at ~500 min in Fig. 4) can be considered as the first nucleation event occurring in the system and allows us to perform a quantitative analysis of the thermodynamics and kinetics of this nucleation event using classical nucleation theory (CNT), e.g. [40]. To this end, the transmittance curves were analysed to estimate the induction times ($t_{ind}$) following the linear fits procedure illustrated in Fig. 3 and Fig. S1 (see Supplementary Material). Before proceeding, it is important to mention here (see discussion above) that the drop in light transmittance used to determine the induction time corresponds to the formation of amorphous spheroids and not to the final C-S-H crystallites. This was previously reported by Krautwurst *et al* [20] and is confirmed in the present study by cryo-TEM analysis (see below). Consequently, a few approximations need to be made to calculate *SI* and to apply the CNT (Eq. 2). First, lacking details on the amorphous spheroids, the solubility product of the final material, identified as β-C-S-H (see Supplementary Material) was used to calculate the *SI* values. Second, the only exact value for a molar volume available in the literature is that corresponding to a nanocrystalline C-S-H phase. Here, we made the choice to use the molecular volume of the amorphous spheroids estimated in the work of Krautwurst [20]. Finally, the use of the master equation of CNT (Eq. 2) implies that the nature of the formed precipitate and the precipitation regime are the same for all the supersaturation conditions. However, data fitted using all the supersaturation range, shows a deviation at high supersaturation values (see Figure S2). This may points to a change in the nature of the precipitate at high Ca and Si concentrations or in the precipitation regime, as it will be discussed below. A common way to fix this issue is to separate the experimental data in two parts: low and high supersaturations. Here, data from the low supersaturation range has been fitted using Eq. 2 considering $C_0$ as a constant independent on the *SI* value, an approximation widely used in the literature.

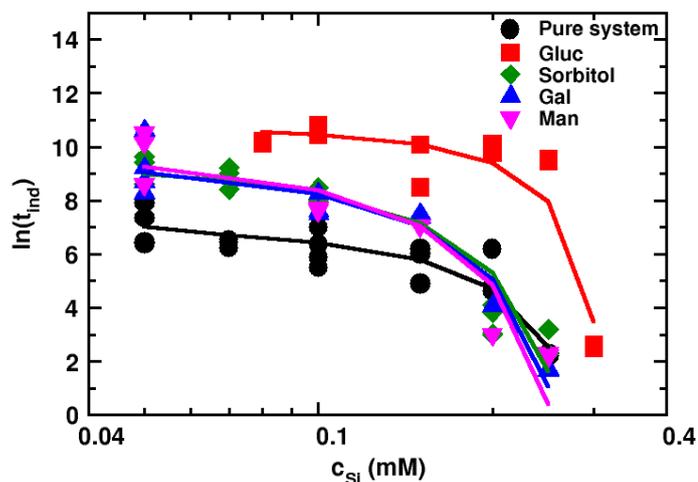

**Figure 5.** Napierian logarithm of the induction time as a function of the initial silicate concentration for the pure system (black) and in presence of 10 mM of organics. The $CaCl_2$ and NaOH concentrations were the same for all experiments, that is 15 mM and 80 mM, respectively. The lines are guides to the eyes.

The variation of the estimated induction times for increasing concentrations of silicate for the four organics-C-S-H systems in comparison with the pure system are shown in Figure 5. As expected, the induction time of nucleation is observed to decrease when the initial silicate concentration is raised for all systems studied. An overall increase of the induction times when organics are present is also observed. The highest values of induction time are found for gluconate, which are two to three orders of magnitude higher than those of the other systems, i.e. without organics and with hexitols. Similar $t_{ind}$ values are observed for all three hexitols, well above those of the pure system but only in the range of low silicate additions. Indeed, in the range of high silicate concentrations, values similar to the pure system are observed with a tendency to be lower. Finally, above a threshold value of silicate concentration the induction time is observed to decrease significantly for all systems.

*3.2. Precipitation regimes of the amorphous C-S-H spheroids*

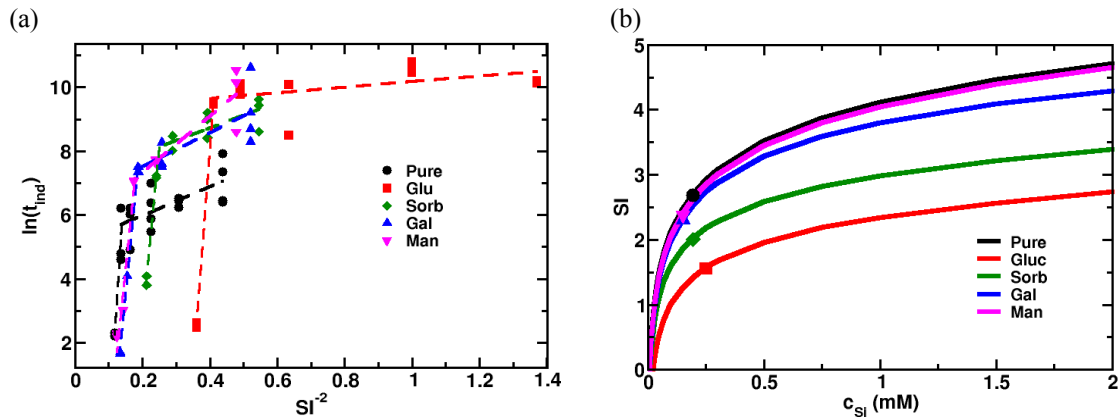

Figure 6: (a) Napierian logarithm of the induction time ($\ln t_{ind}$) of precipitation as a function of the reverse of the saturation index square ($SI^{-2}$); (b) Calculated saturation index ($SI$) with respect to C-S-H versus silicate concentration of the metastable solution; in the pure system and in presence of the organics. The SI values have been calculated taking into account the complex formation of organics with Ca, Si and OH, see § 2.3. The organics and salt concentrations are the same as in Fig. 5. The symbols in (b) indicate the position at which the transition of precipitation regime is observed in (a). One can note that this transition is found to be close to the point of maximum curvature. The $SI$ values at which this transition occurs decrease in the order pure > Man ~ Gal > Sorb > Glu.

From the initial Si, Ca and OH concentrations of the metastable solutions the saturation indexes with respect to β-C-S-H, accounting for the formation of the Si-Ca-OH-organic complexes, can be determined (see § 2.3). This allows us to plot the induction times shown in Fig. 5 as a function of $SI^{-2}$. The typical $\ln(t_{ind}) = f(SI^{-2})$ plots are given in Fig. 6-a and the corresponding variation of $SI$ versus the silicate concentration in the metastable solution are shown in Fig. 6-b. As can be seen, the change in the $SI$ for the different systems at the same total silicate concentration (Fig. 6-b), explains to some extent the differences in the measured induction times among the organics and between the organics and the pure system (Fig. 6-a). This is particularly true in the low supersaturation range (high $SI^{-2}$) when comparing the gap between the gluconate and the hexitol curves in Fig. 5 and 6. In other words, the change in the kinetics of precipitation induced by the organics is partly explained by the solution chemistry, i.e. organics-Ca-Si-OH complex formation. Also, the sudden increase in precipitation rate observed in Fig. 5 and, in particular, the critical point beyond which this occurs is now more clearly distinguishable. What is more, the $SI$ value at which this transition is observed is found to decrease in the order pure > Man ~ Gal > Sorb > Gluc, see Fig. 6.

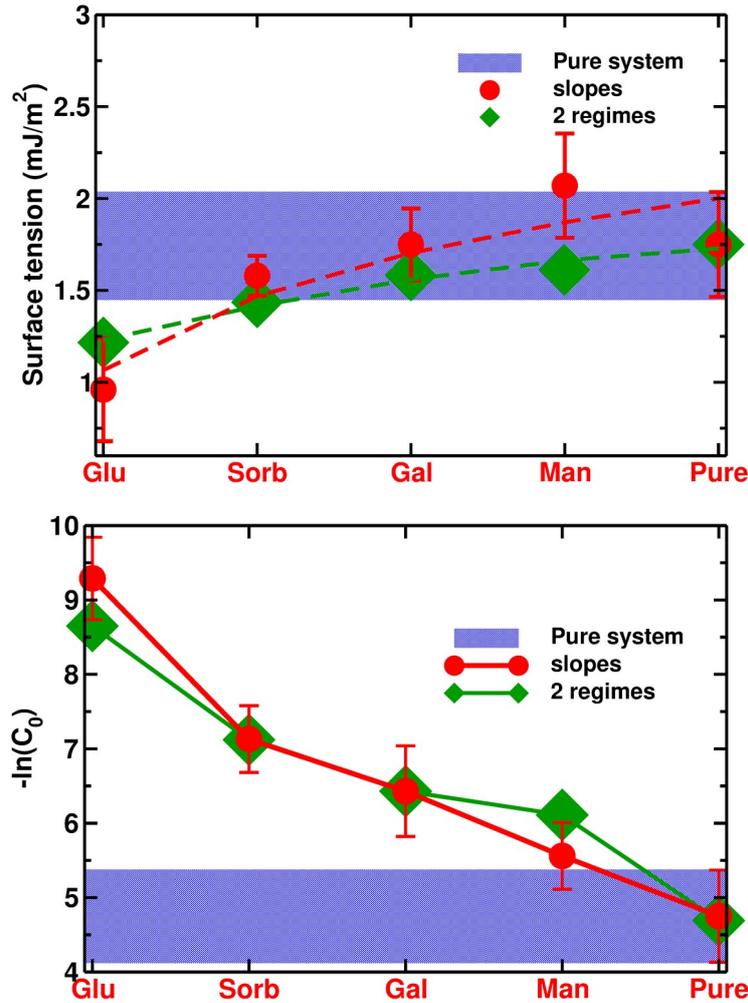

Figure 7: Top: interfacial free energies of the amorphous particles as obtained for the pure and organic-containing systems. Bottom: Corresponding values of the nucleation kinetic pre-factor, $C_0$. The organics and salt concentrations are the same as in Fig. 5. *Red symbols*: $\gamma$ and $C_0$ values obtained by linear fit with the CNT (Eq. 2) of the slow precipitation regime (high $SI^{-2}$ range) shown in Fig. 6. *Green symbols*: $\gamma$ and $C_0$ values obtained with Eqs. 5 and 6 which assumes that the fast precipitation regime at high saturation indexes (low $SI^{-2}$) is due to a gradual transition to a spinodal decomposition as a result of a negligible free energy barrier to the nucleation, $\Delta G < 1kT$, see the main text. The dotted lines are guides to the eyes.

To a good approximation, the $\ln(t_{ind}) = f(SI^{-2})$ curves can be described by two straight lines, see Fig. 6-a, compatible with the existence of two distinct precipitation regimes delimited by two rather well defined supersaturation regions. In the following, they will be referred to as slow and fast precipitation regimes. Similar observations and descriptions of the precipitation kinetics were obtained for the homogeneous nucleation of C-S-H without organics, see e.g. [41], as well as of many other inorganic systems, see e.g. [42,43]. The change in the precipitation kinetics/regimes was often interpreted as a transition from homogeneous (at high *SI*) to heterogeneous nucleation (at low *SI*), see e.g. [41,43]. In the case of C-S-H however, the work of Kraustwurst demonstrated that in the low *SI* range (high $SI^{-2}$) the precipitation of C-S-H is controlled by homogeneous nucleation. As we will argue below, the observed transition at high *SI* (low $SI^{-2}$) is instead suggested here to be related to a change in the precipitation mechanism toward a *spinodal decomposition,* where the work of nucleation ($\Delta G$) is *suppressed*. The proposed mechanism is inspired by the theoretical work of Binder, see e.g. [44] and [45], and recent simulation works [46-49].

*3.3. Thermodynamics and kinetics of formation in the slow precipitation regime*

In the slow precipitation regime, low *SI* (high $SI^{-2}$) range in Fig. 6, a data analysis based on the classical nucleation theory (Eq. 2) allows obtaining values of interfacial free energy or surface tension, $\gamma$ and the kinetic pre-factor, $C_0$, from the slope of the data and the y-intercept in Fig. 6a. The results are given by the red points and curves in Fig. 7. The interfacial free energies are calculated with the molecular volume of the amorphous particles estimated from molecular simulations to be ~19 times greater than that of C-S-H (~$6^3$ $\text{A}^3$), see [20]. This in part explains the very low values of the interfacial free energies reported in Fig. 7a. These fall in the range between 0.75 and 2.25 mJ/m$^2$. Without taking this change in volume into account, we obtain values that are an order of magnitude higher, ~13 mJ/m2 for the pure system. As it is expected, the adsorption/incorporation of organics seems to decrease the interfacial free energy of the amorphous particles. However, except for gluconate, which yields a significantly lower $\gamma$ value than the pure system, no assertions can be made about the hexitol effect due to the large uncertainties on $t_{ind}$ and $\gamma$ data. The situation changes for the values of the kinetic pre-factor, $C_0$, for which a clear distinction between the systems can be made, see Fig. 7b. In other words, the effect of polymers on the interfacial free energy is much lower than on the kinetic pre-factor. The gluconate system shows the lowest values of $C_0$ (highest -ln($C_0$)) followed in increasing order by sorbitol, galactitol and mannitol. Except for mannitol they all yield a lower value of $C_0$ than the pure system. Interestingly, this order matches that found for the complexation constants of $Ca^{2+}$ and silicates with these organics: gluconate was found to have the strongest ability to form polynuclear complexes with $Ca^{2+}$ and silicates, followed by sorbitol, which showed a weaker complexation level, and then by galactitol and mannitol, with the weakest tendency to form complexes. Similarly, gluconate is also the organic molecule that yields the lowest interfacial free energy and was found to adsorb the most on C-S-H [38]. When the concentration of the organics is decreased to 1 mM, or even to 0.1 mM, the same behaviour is observed but the decrease of the kinetic prefactor and the interfacial free energy with respect to the pure system becomes weaker when less organics is present in the system (see data in the Supplementary Material).

*3.4. Formation mechanism in the fast precipitation regime*

In the fast precipitation regime at high *SI* (low $SI^{-2}$ range, in Fig. 6), the data can arguably be ascribed to a mechanism of *spinodal nucleation,* which can be considered as a *gradual transition from "classical" nucleation to a spinodal decomposition* and where *ΔG* is *substantially reduced*. This mechanism was studied in detail with simulations in the 90's by Binder to explain the peculiar phase transition observed for polymers when approaching the spinodal [50-52], that is where the nucleation barrier is virtually null and the system becomes unstable. But, the theory of the transition from nucleation to spinodal decomposition is accurate only in the case of infinitely long range and weak interactions between species. In this limit, the transition is found to be singular (abrupt). When the range of interactions is finite (as in reality), Binder showed, in agreement with experiments that the transition from "classical" nucleation to spinodal decomposition is *gradual* and the width of this gradual transition (range of supersaturations) depends on the range of interactions among the species. Binder further found that the onset of the *gradual transition to the spinodal decomposition* occurs for a threshold value of the nucleation free energy barrier not exceeding a few times the kinetic energy of the system, typically for $1\ kT \leq \Delta G^* \leq 10\ kT$. In recent years, the transition to the spinodal decomposition has regained attention in the context of non-classical nucleation as it is regarded as a plausible explanation for the non-classical nucleation pathway observed for several inorganic and organic solid phases, see e.g. [22-27,29,47-49].

Different observations support the fact that the observed transition to the fast precipitation regime can be explained as a gradual transition from nucleation to spinodal decomposition. First, the whole data set, including low and fast precipitation regimes (Fig. 6), cannot be correctly fitted using the equation of the classical nucleation theory (see Fig. S2). Second, the order of the slopes in the fast and slow precipitation regimes is not consistent. In particular, the observed slope order in the fast precipitation regime, pure > gluconate > sorbitol > galactitol > mannitol, is incompatible with homogeneous nucleation as it would violate the basics concepts of Gibbs thermodynamic on adsorption and interfacial free energies. Third, the onset of the fast precipitation regime is observed to happen just before the flattening of the *SI* versus $c_{Si}$ curves (Figure 6b). Actually, a stabilization of *SI*, equivalent of the driving force, when increasing the solute concentration is expected when approaching the spinodal [47]. Thermodynamic definition of the spinodal gives, indeed, $\partial SI/\partial c = 0$. Fourth, not too high values of *SI* are expected at/close to the spinodal because of the very small values of the interfacial free energies as obtained for the amorphous spheroids because $\Delta G \propto \gamma^3$. Finally, below we show that the threshold value for the nucleation barrier at the observed transition of precipitation regimes, $\Delta G^*$, is very small, well in accord with a gradual transition to a spinodal decomposition.

The lack of a theory to describe the transition of the precipitation regime can be circumvented by using the fact that for $\Delta G \geq \Delta G^*$, that is below a threshold Saturation Index value, $SI^*$, the precipitation is driven by nucleation *alone*, where classical nucleation theory can be applied, see above discussion. Consequently, the interfacial free energy and the kinetic prefactor, but also $\Delta G,^*$ can be evaluated from the threshold values of $t_{ind}^*$ and $SI^*$ marking the end of the "classical" nucleation regime and the onset of the transition, see Fig. 6a. Indeed, within the CNT it is straightforward to show that *at the onset of the transition* the following relationships apply for the interfacial free energy

$$\frac{\gamma_i}{\gamma_{ref}} = \left(\frac{SI_i}{SI_{ref}}\right)^{2/3} \tag{3}$$

and the kinetic prefactor

$$(\ln t_{ind})_i - (\ln t_{ind})_{ref} = (\ln C_0)_{ref} - (\ln C_0)_i \tag{4}$$

Note that Eqs 2-3 assumes that the molecular volume of the precipitated phase and $\Delta G^*$ are the same for systems *i* and *ref*. What is more, when $\Delta G^*$ is known, absolute values of $\gamma$ and $C_0$ can be obtained from $t_{ind}^*$ and $SI^*$ (and vice versa) with the following equations,

$$\gamma = \left(\frac{\Delta G^*}{kT}\frac{3(\ln 10\, SI^*)^2}{16\pi v_m^2}\right)^{(1/3)} \tag{5}$$

$$\ln t_{ind}^* = \frac{\Delta G^*}{kT} - \ln C_0 \tag{6}$$

Fig. 7 provides a comparison of $\gamma$ and $C_0$ values obtained from the slow precipitation regime (with Eq. 2) with those (green diamonds and lines) determined from the onset of the fast precipitation regime using Eqs. 5 and 6 assuming that the corresponding nucleation free energy barrier is identical for all systems and equal to the kinetic energy, that is $\Delta G^* = 1\ kT$. The agreement is remarkably good and also extends to lower organics concentrations (See Figs. S4 and S4 in the Supplementary Material). Interestingly, the shift of the transition point between the two precipitation regimes to lower supersaturations (higher $SI^{-2}$) observed in Fig. 6, in the order pure < Man < Gal < Sorb < Glu, can thus be explained by a decrease in the interfacial free energy in the presence of the organics, see Eq. 3. The excellent agreement we obtained for a large number of systems, along with the small and equal $\Delta G^*$, strongly supports our hypothesis that the high end of $SI$ ($SI > SI^*$) experiences fast precipitation as a result of a gradual transition from nucleation to spinodal decomposition, due to $\Delta G$ being less than 1 $kT$. At lower $SI$ values ($SI \leq SI^*$), the slow precipitation is driven by homogeneous nucleation, with $\Delta G$ being greater than 1 kT.

### 3.5. Physical characterization of the precipitates

Cryo-TEM images of the precipitates formed in the presence of 1 mM gluconate and 5 mM sorbitol, imaged at different time points of the nucleation process, are shown in Figure 8. Particles show a spherical morphology in all cases, with larger particles (~500 nm to 1 μm) formed in the presence of gluconate. The size of the particles formed in the presence of sorbitol is considerably smaller, ~10-50 nm, similar to the sizes found by Krautwurst et al. for the pure system under identical chemical conditions of solution supersaturation (18). The electron diffraction pattern shown as inset in Fig. 8b reveal that the formed particles are amorphous.

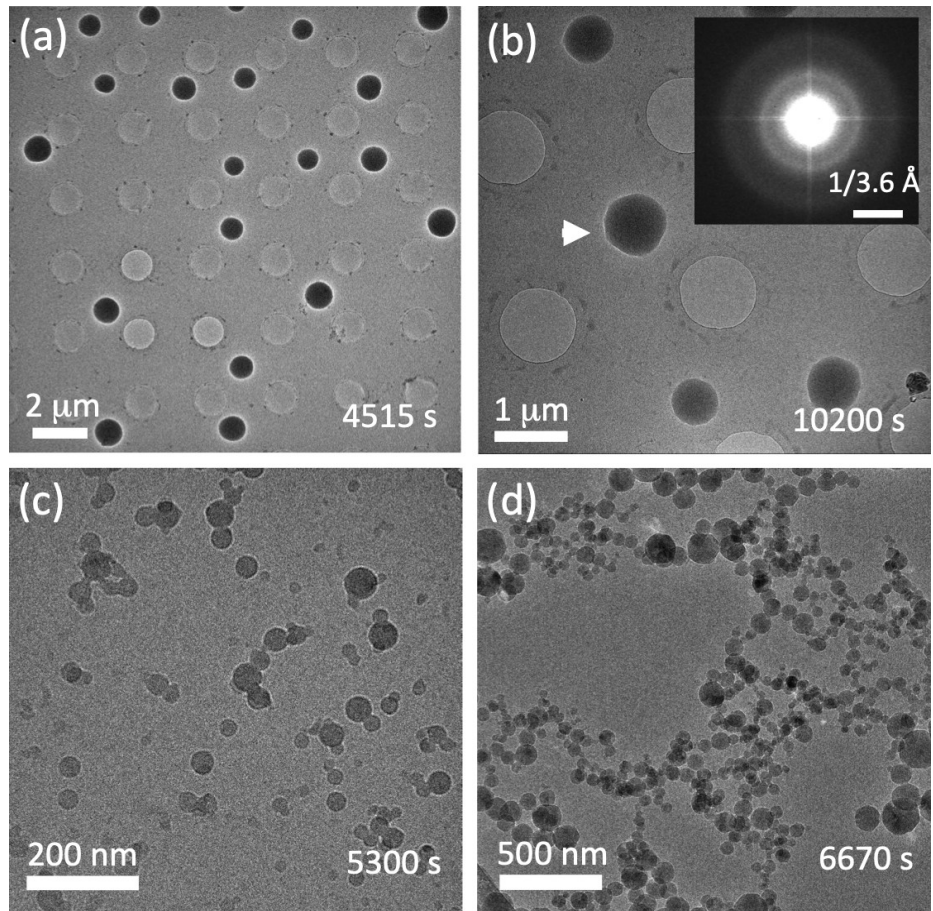

**Figure 8.** Cryogenic TEM images of the early stages of C-S-H precipitation in the presence of 1mM gluconate (a,b) and 5 mM Sorbitol (c,d). White arrow in (b) indicates the particle that was used for selected area diffraction (SAED) shown in the inset of (b). Times elapsed between the onset of the reaction and the cryo-quenching of an aliquot of the reacting solution.

SAXS data for some selected times are represented in Fig. 9. In all cases, an increase of the intensity is observed over time. However, the pure system exhibits a clear induction time, with the intensity increasing only after 10 min, whereas no induction time is observed for the gluconate-containing systems (see evolution of the SAXS invariant in Figure 9). This is in agreement with the turbidity data shown before, where the formation of the complexes with gluconate induced an immediate decrease of the transmittance upon mixing of the silicate and gluconate-bearing calcium solutions.

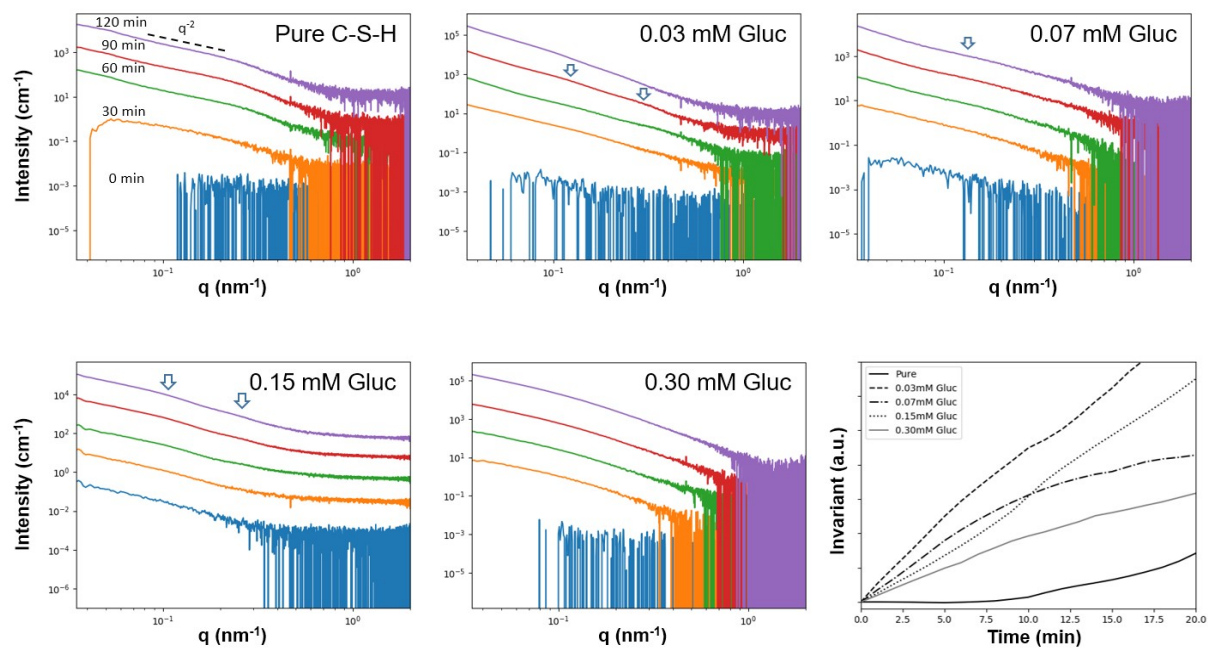

**Figure 9.** SAXS patterns and evolution of the SAXS invariant over time for the pure and gluconate-bearing systems. The arrows indicate bumps in the intensity coming from particle scattering within the precipitates. (the colour code for time is the same in all graphics)

SAXS patterns from the pure system evolve from a particle scattering characteristic of a very polydisperse system (30 min), with no characteristic form factor, to a system where the intensity distribution follows a $q^{-2}$ dependence, as expected for plate-like systems. It is important to keep in mind that SAXS is sensitive to density variations, and that these can also take place within the amorphous spheroids, which are very hydrated systems. A change of slope is observed at $\sim 3 \cdot 10^{-1}$ nm$^{-1}$, which is tentatively ascribed to the thickness of the plates, which would be in the order of 20 nm. However, the polydispersity of the data precludes a detailed analysis of the morphology. It is interesting to observe that the plate-like particles are observed after ~60 mins, but, according to Krautwurst et al., the particles are still amorphous at this time. Our hypothesis is that silicate molecules and calcium atoms within the amorphous spheroids (as identified as silicate dimers by Krautwurst et al.) adopt plate-like atomic arrangements, which occur as a preliminary step in the crystallization process. The low concentrations used in the experiments resulted in the absence of a signal at any point in the wide-angle detector that was available at the SAXS beamline.

Data from the systems containing gluconate are more complex. Different small bumps (see arrows in Fig. 9) are observed, which could indicate the presence of particles being aggregated into large networks. Indeed, for almost all cases (except for the system with 0.30 mM gluconate) the low $q$ part of the data follow a power law that can be fitted using a $I \propto q^{-n}$ law (Fig. S3). The results show that the values of the exponent tend towards values higher than 2 in almost all cases, which are typically associated to mass fractals aggregates, tending to surface fractal when the values are higher than 3 (see Fig. S3). The case of the system with 0.30 mM of gluconate is different from the rest: all the data exhibit a certain degree of curvature, which again can be explained by a very polydisperse system with no particular form factor, as it was the case for the early stages (~30 min) of nucleation of the pure system.

Another observation that is worth mentioning is the fact that the background signal is significantly higher in the system with 0.15 mM gluconate than in the rest. This could point to the presence of sub-nm entities that were present in the solution (ion pairs, small complexes, etc.). Although it is not possible to determine the exact nature of these from the data, they seem to support the early decrease in light transmittance over time in the presence of gluconate and the corresponding discussion on the formation of organics-Si-Ca-OH complexes. Overall, the cryo-TEM and scattering data further confirm that organics do not alter the mechanism of the nucleation pathway, but rather slow down the different processes taking place en route from dissolved species to the final solid phase.

## 4. Discussion & Conclusions

It still needs to be determined to what extent the obtained data can enhance our understanding of the homogeneous nucleation of C-S-H and the influence of small organic molecules. Without organics, the present study confirms the finding of Krautwurst [20], i.e. C-S-H crystallization takes place in two main stages, starting with the nucleation of well-dispersed amorphous particles followed by their aggregation and concomitant re-organization into the final C-S-H crystallites. It should be stressed here, in light of previous works on C-S-H formation, that the nucleation of the amorphous spheroids is initiated by the rapid formation of silicate oligomers in the supersaturated solution, as revealed by e.g. the mass spectroscopy analysis of metastable solutions at different times ([53,54]), and triggered by the $Ca^{2+}$ mediated coagulation of these oligomers. The degree of polymerization of the silicate oligomers is unfortunately not well characterized but seems to depend on the pH and calcium concentration of the supersaturated solution, which also controls the final C/S of the C-S-H. Indeed, partial and indirect literature data obtained from the analysis of the amorphous C-S-H spheroids seem to indicate the presence of relatively long silicate oligomers at low pH (pH << 12) [54] and of silicate dimers at high pH (pH >> 12) [20]. In the present study, and contrary to the case of systems with organics, the silicate oligomers did not significantly decrease the light transmittance. This may confirm that silicate oligomers are short in these high pH conditions [e.g. 20]. Interestingly, our SAXS data appears to suggest that silicate molecules and calcium atoms within the amorphous spheroids adopt plate-like atomic arrangements prior to the crystallization of the final C-S-H particles. This observation is in line with simulations that will be the topic of a forthcoming paper. Finally, our kinetic analysis of the first stage of the homogeneous formation of C-S-H clearly indicates the existence of two regimes: a slow precipitation regime at low and moderate supersaturations and a fast precipitation regime at high supersaturations. In the range of low and moderate *SI*, the precipitation of the amorphous spheroids can be described by the classical nucleation theory. Above a threshold value of *SI*, the formation rate of the amorphous spheroids is found to increase dramatically as a result of a negligible free energy barrier to the nucleation, estimated to be $\Delta G < \Delta G^* = 1\ kT$. A similar value for the free energy barrier at the transition, $\Delta G^*$, is obtained for all four organics at the three concentrations studied; 0.1, 1 and 10 mM of organics. This low value of $\Delta G^*$ for a large number of C-S-H systems provides strong evidence that the precipitation mechanism of the amorphous C-S-H spheroids in the high *SI* range is a *gradual transition from nucleation to spinodal decomposition*. This process, referred to as *spinodal nucleation* by Binder [45], is still an activated process, i.e. $\Delta G > 0$, but it shares characteristics of spinodal decomposition, such as the formation of ramified clusters that merge and grow over time. This key result is summarized in the master curve shown in Fig. 10, which was obtained through the application of Eqs. 5 and 6.

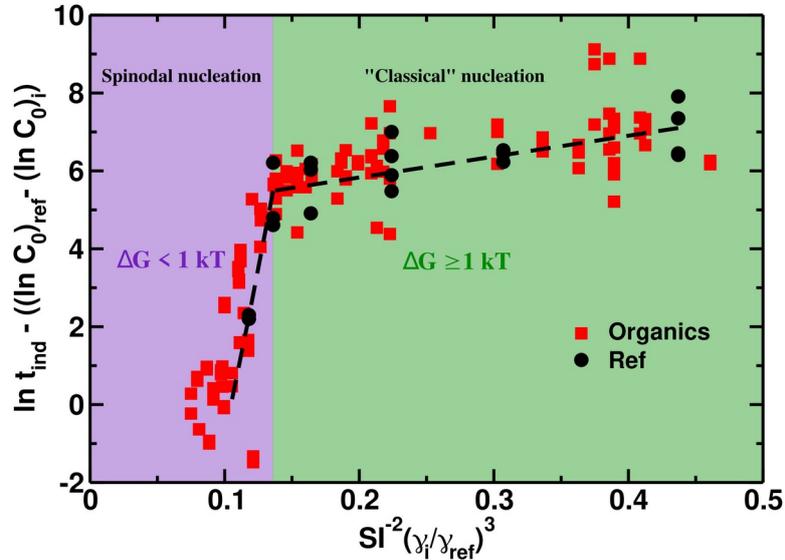

Figure 10: Induction time of the amorphous C-S-H spheroid formation versus saturation index for all systems (Fig. 6-a and Fig. S4) re-scaled according to Eqs. 3 and 4 using the pure system as the reference (ref). The black dashed line serves as guide to the eye.

Our data also show that, when the supersaturation with respect to C-S-H is moderate, organics do not alter the mechanism of the C-S-H nucleation pathway, but rather slow down the different processes taking place en route from dissolved species to the final solid phase. The formation of organic-Si-Ca-OH complexes plays a role in explaining why the driving force for nucleation (Saturation Index) decreases, especially in the presence of sodium gluconate, as shown in Figs. 5 and 6b. However, the differences between the various organic systems and the pure system cannot be fully explained by solution chemistry, as demonstrated in Fig. 6a, which compared the induction times for different systems at the same $SI$. The CNT analysis of the slow precipitation regime reveals that the slow down of nucleation in the presence of the organics is mostly noted in the kinetic prefactor $C_0$ of the nucleation rate, $J_n = C_0 \cdot \exp(-\Delta G/kT)$, which significantly decreases in the following order: pure ~ mannitol > galactitol > sorbitol >> gluconate. This order turns out to be inversely proportional to the power of the molecules to form organics-Si-Ca-OH complexes. Incidentally, this complexation power seems also to be correlated to the initial decrease in the light transmittance with organics, observed to be larger for gluconate than sorbitol, as well as the height of the background signal of the SAXS data with gluconate. The light transmittance and SAXS data in the supersaturated solution prior to the nucleation of the amorphous spheroids are consistent with the hypothetical formation of clusters of organic-Si-Ca-OH complexes. Based on these observations, it is reasonable to suggest that the decrease in the kinetic prefactor in the presence of organics is due to the stabilization or reduced diffusion coefficient of the Ca-OH-silicate-oligomer clusters caused by organics complexation. It is noteworthy that the case of gluconate does not show a second drop in the light transmittance (as seen in Fig. 2). This could indicate that the presence of organics stabilizes the amorphous spheroids against aggregation and concomitant crystallization. This type of effect, in which an organic additive modifies crystal kinetics by imposing steric barriers to aggregation, has previously been reported for other systems such as $CaSO_4$ [55]. However, additional data and complementary experiments are required to confirm these hypotheses.

When the solution becomes unstable at high $SI$ ($SI > SI^*$) the precipitation rate of the amorphous C-S-H spheroids in the presence of organics appears to be faster that in the pure system. What is more, there is a reversal in the order of precipitation rates (gluconate > sorbitol > galactitol > mannitol > pure) as compared to the classical nucleation regime for $SI < SI^*$ (pure > mannitol > galactitol > sorbitol > gluconate), see Fig. 6-a. This trend can be attributed to the decrease of the interfacial free energy of the amorphous spheroids due to the complexation/adsorption of the organics (Eq. 3), as shown by the slope analysis in the "classical" nucleation regime. However, the uncertainty associated with the data prevents any conclusions regarding the effect of hexitol, see Fig. 7a.

Our data thus strongly suggest that the interfacial free energy of the amorphous spheroids decreases in the following order: pure > mannitol ~ galactitol > sorbitol > gluconate, in line with the affinity of the same organics with C-S-H [38]. Noteworthy, application of Eqs. 3 and 4 leads to a master curve in the "classical" nucleation regime as shown in Fig. 10. This also reveals that, although the change in the kinetic prefactor $C_0$ is the most marked, a small change in interfacial free energy cannot be disregarded. One clear consequence is the observed shift of $SI^*$ values when organics are added (Fig. 6-a), which is found to be proportional to the cubic power of the surface tension (Eq. 3). We also note that, for the same reasons, small variations in the surface tension can have large consequences on the rate of the "classical" nucleation in the low *SI* range, as shown in Fig. 11. Actually, these are here blurred by the large scattering in the data of the measured induction times. A sensitivity analysis of the effect of the interfacial free energy is shown in Fig. 11. Values of the height of the nucleation barrier –normalized by the thermal energy, $k_BT$- are given in the colour bar, and are plotted as a function of the interfacial free energy (x axis) and the saturation index of the solution (y axis). The range of values in both x and y axes have been set to values that are used or found in the experiments reported here. The results show that the nucleation barrier can vary between very small values (< 5) where nucleation can happen spontaneously, to values in the order of ~10 at which induction times for nucleation tend to infinite. This sensitivity analysis leads us thus to make a cautionary note about the results of the interfacial free energies: the relatively large error bars precludes any detailed discussion about the implications of the values found in terms of the height of the nucleation barrier. Finally, we remark that this large uncertainty is due to the silicate, here present in very small amounts (see § 2), indicating that the silicate is the limiting species of the C-S-H nucleation.

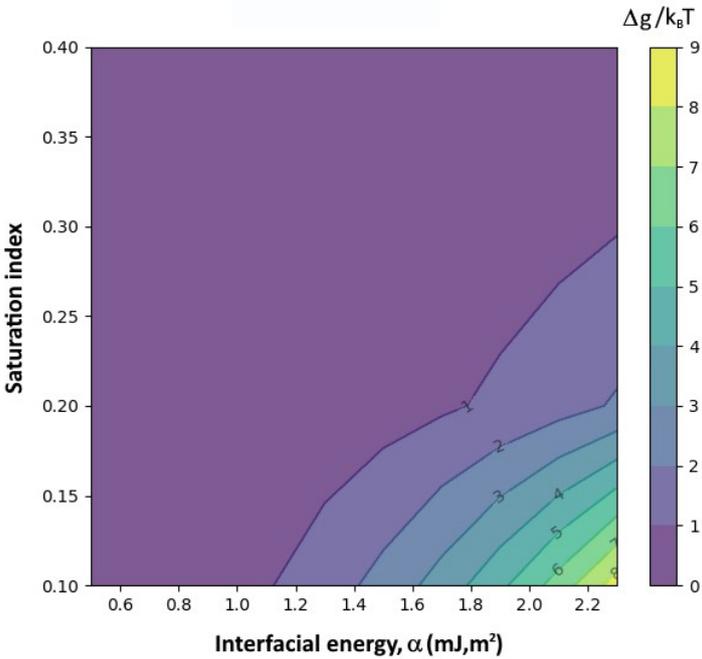

**Figure 11.** Height of the nucleation barrier (in $k_BT$ units) as a function of the interfacial energy and saturation index of the solution with respect to β-C-S-H.

Based on the results obtained, some (preliminary) conclusions can be drawn about the impact of the findings on our understanding of the hydration process of $C_3S$. Firstly, the results definitively demonstrate that the hydration delay caused by hexitols is a result of a slowing down of the precipitation of C-S-H. While this was suggested by previous findings by Nalet et al. [9,11,16] , this has now been clearly established. Secondly, our results strongly suggest that this delay in precipitation is mainly of kinetic origin ($C_0$). We can reasonably hypothesize that this is explained by the much slower diffusion of the reactive species (Ca, Si and OH$^-$) due to their complexation with the organics. Finally, we could show that at high silicate concentration, i.e. high *SI* range, the precipitation is as fast or faster than in the pure case. The important question is then what is the consequence of the rapid formation of amorphous C-S-H spheroids on the dissolution of $C_3S$? The case of gluconate is even more complex since it was also found to decrease the dissolution rate of $C_3S$ [19]. Whether C-S-H precipitation or $C_3S$ dissolution explains the retardation of $C_3S$ hydration in the presence of gluconate remains an open question.  These are relevant questions that we leave for future studies.

To summarize, the results presented in this work confirmed the previous observations made by Krautwurst et al [20] regarding the nucleation of amorphous spheroids and the subsequent crystallization of these into crystalline C-S-H particles, as evidenced by the characteristic drops in transmittance. The same steps were also observed in systems containing organic molecules. Cryo-TEM observations confirmed that a nucleation pathway via an amorphous precursor takes place also in the presence of gluconate and sorbitol. The analysis of the induction times of the amorphous C-S-H spheroid formation versus saturation index reveals two distinct precipitation regimes, a "classical" nucleation regime at low and moderate saturation indexes where $\Delta G > 1$ kT and a spinodal nucleation regime [44] at high *SI,* which can be understood as a gradual transition from nucleation to spinodal decomposition, where $\Delta G < 1$kT. The slope analysis of the induction time in the nucleation regime yields similar values for the interfacial energies, considering the uncertainties, except for gluconate whose obtained values that were clearly lower than those of the other systems. This points to a regime principally controlled by the kinetic prefactor, though the results have to be taken with caution given the large error bars and the sensitivity of the induction times to the value of the interfacial energy. The analysis of the kinetic factor, which includes terms related to the physics of the collisions between clusters formed during the nucleation process, yields an interesting observation: the values of this factor increase from the pure system to the gluconate-containing systems, and are inversely proportional to the values of the complexation constants found for each of the organics with the C-S-H containing species [38]. This observation may suggest that the mechanism by which the organics delay the nucleation is related to the stabilization of the pre-nucleation clusters of silicate and calcium by organics, slowing down the formation of the metastable amorphous spheroids. Finally, in the high *SI* range, a reversal in the order of precipitation rates (gluconate > sorbitol > galactitol > mannitol > pure) as compared to the nucleation regime (pure > mannitol > galactitol > sorbitol > gluconate) is observed. This is explained by the decrease of the interfacial free energy of the amorphous spheroids due to organics sorption/complexation.

In spite of the interesting observations found here, it is unknown whether the amorphous spheroids are a requirement for C-S-H formation in real cement pastes. Experimental studies highlighted that C-S-H formation can be described as primary heterogeneous nucleation and growth [56,57] under real cement conditions. Indeed, given that dislocations, impurities or grain boundaries are already present in the primary materials, the nucleation process could occur more easily close to solid-liquid interfaces. Future experiments addressing this issue –multi-step pathways of heterogeneous nucleation- should be designed to provide a more complete view of this complex system.

## Acknowledgments

We would like to thank for the financial support from Nanocem (core project 15). Marc Malfois from the synchrotron ALBA and Karen Alloncle from ICB-Dijon are greatly acknowledged from their respective support in SAXS and induction time measurements. We are also very grateful for many helpful discussions with the representatives of the industrial partners L. Pegado, J.H. Cheung, V. Kocaba, P. Juilland and M. Mosquet and their interest.

# Supplementary Material for:

# Mechanisms and kinetics of C-S-H nucleation approaching the spinodal line: Insights into the role of organics additives

Christophe Labbez[1,*], Lina Bouzouaid[1], Alexander E.S. Van Driessche[2], Wai Li Ling[3], Juan Carlos Martinez[4], Barbara Lothenbach[5], Alejandro Fernandez-Martinez[2,*]

[1] ICB, UMR 6303 CNRS, Univ. Bourgogne Franche-Comté, FR-21000 Dijon, France
[2] Univ. Grenoble Alpes, Univ. Savoie Mont Blanc, CNRS, IRD, IFSTTAR, ISTerre, FR-38000 Grenoble, France
[3] Univ. Grenoble Alpes, CEA, CNRS, IRIG, IBS, FR-38000 Grenoble, France.
[4] Non-Crystalline Diffraction Beamline−Experiments Division, ALBA Synchrotron Light Source, Cerdanyola del Vallès, SP-08290 Barcelona, Spain
[5] Empa, Concrete & Asphalt Laboratory, CH-8600 Dubendorf, Switzerland

**Table S1.A.** Concentrations of the different complexes formed in the supersaturated solutions, obtained from the PHREEQC simulations, using a gluconate concentration of 10mM. The concentration of calcium was fixed at 15mM. Silicate concentration values were fixed at 0.08mM, 0.10mM, 0.15mM and 0.20mM.

| pH | Gl | Ca | Si | $Ca^{2+}$ | $CaOH^+$ | GlCaOH | $Gl_2Ca_2OH_2^{2-}$ | $Gl_2Ca_3(OH)_4$ | $GlCa^+$ | $Ca_3Gl_2(H_3SiO_4)_2(OH)_2$ |
|---|---|---|---|---|---|---|---|---|---|---|
| 12.75 | 10 | 15 | 0.08 | 64.64 | 1.75 | 1.84 | 0.05 | 2.14 | 0.21 | 0.012 |
| pH | Gl | Ca | Si | $Ca^{2+}$ | $CaOH^+$ | GlCaOH | $Gl_2Ca_2OH_2^{2-}$ | $Gl_2Ca_3(OH)_4$ | $CaGl^+$ | $Ca_3Gl_2(H_3SiO_4)_2(OH)_2$ |
| 12.75 | 10 | 15 | 0.10 | 4.64 | 1.75 | 1.84 | 0.05 | 2.13 | 0.21 | 0.017 |
| pH | Gl | Ca | Si | $Ca^{2+}$ | $CaOH^+$ | GlCaOH | $Gl_2Ca_2OH_2^{2-}$ | $Gl_2Ca_3(OH)_4$ | $CaGl^+$ | $Ca_3Gl_2(H_3SiO_4)_2(OH)_2$ |
| 12.75 | 10 | 15 | 0.15 | 4.63 | 1.75 | 1.83 | 0.05 | 2.12 | 0.21 | 0.031 |
| pH | Gl | Ca | Si | $Ca^{2+}$ | $CaOH^+$ | GlCaOH | $Gl_2Ca_2OH_2^{2-}$ | $Gl_2Ca_3(OH)_4$ | $CaGl^+$ | $Ca_3Gl_2(H_3SiO_4)_2(OH)_2$ |
| 12.75 | 10 | 15 | 0.20 | 4.63 | 1.75 | 1.83 | 0.05 | 2.11 | 0.21 | 0.046 |

**Table S1.B** Concentrations of the different complexes formed in the supersaturated solutions, obtained from the PHREEQC simulations. Simulations include experiments with sorbitol and sodium gluconate at a fixed concentration of 10mM. Calcium and silicate concentration values were fixed at 15mM and 0.15mM respectively.

| pH | Sor | Ca | Si | $Ca^{2+}$ | $CaOH^+$ | $SorCaOH^+$ | $Sor_2Ca_2(OH)_4$ | $Ca_2Sor_2H_2SiO_4(OH)_4^{2-}$ | $SorCa^{2+}$ | $Ca_3Sor_2(H_3SiO_4)_2(OH)_2$ |
|---|---|---|---|---|---|---|---|---|---|---|
| 12.70 | 10 | 15 | 0.15 | 9.448 | 3.88 | 1.84 | 0.04 | 0.01 | 0.11 | 0.028 |
| pH | Gl | Ca | Si | $Ca^{2+}$ | $CaOH^+$ | GlCaOH | $Gl_2Ca_2OH_2^{2-}$ | $Gl_2Ca_3(OH)_4$ | $CaGl^+$ | $Ca_3Gl_2(H_3SiO_4)_2(OH)_2$ |
| 12.75 | 10 | 15 | 0.15 | 4.63 | 1.75 | 1.83 | 0.05 | 2.11 | 0.21 | 0.046 |

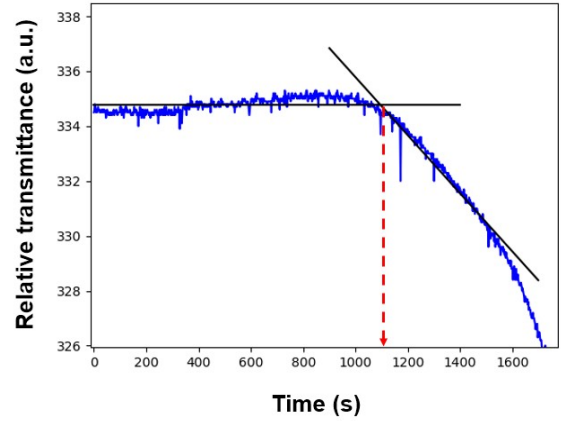

**Figure S1.** Method to determine the induction time from turbidity data, exemplified with the data from the pure system at [Si]=0.10mM. Two linear fits are performed to determine the point at which the transmittance of the solution starts to decay (indicated in red).

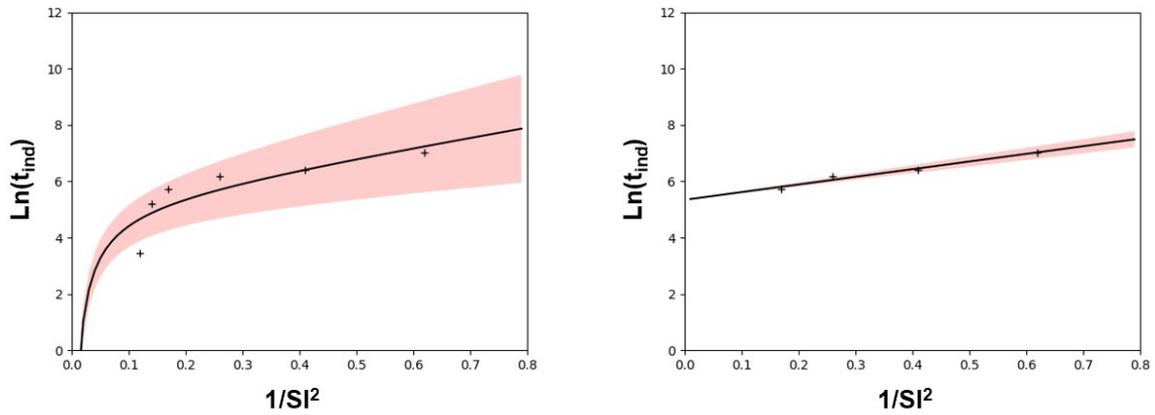

**Figure S2**. Left: Average values of the induction times for the pure system, together with a fit of the classical nucleation theory expression for the induction time (see eq. 1 in the main text). The equation includes a dependence of the kinetic pre-factor ($C_0$) with the supersaturation, which gives rise to the curvature [1]. Right: Fit of the expression with $C_0$ being a constant, and using a fitting range that include only the points at lower supersaturations (i.e., high $1/SI^2$ values).

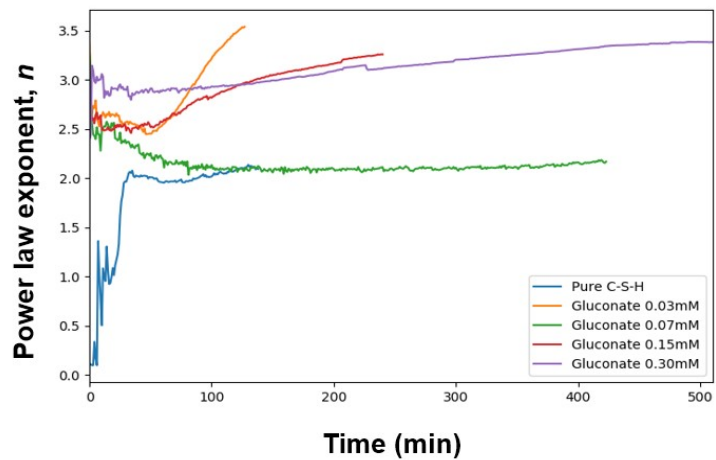

**Figure S3.** Values of the power law exponent used to fit low q part of the data using a $I = q^{-n}$ law.

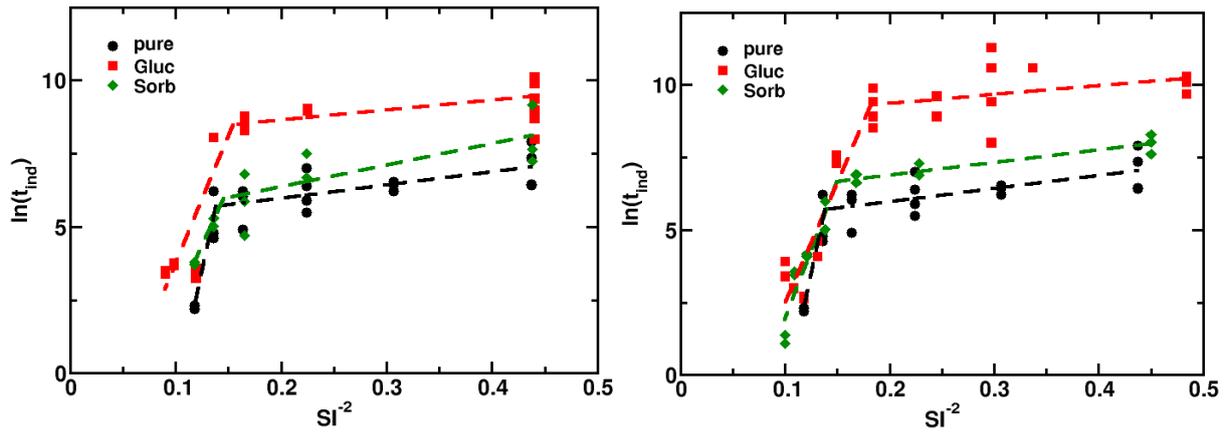

**Figure S4.** Napierian logarithm of the induction time (ln $t_{ind}$) of C-S-H precipitation as a function of the reverse of the saturation index square ($SI^{-2}$) at two different concentrations of organics and in comparison with the pure system; Left: 0.1 mM; Right: 1 mM.

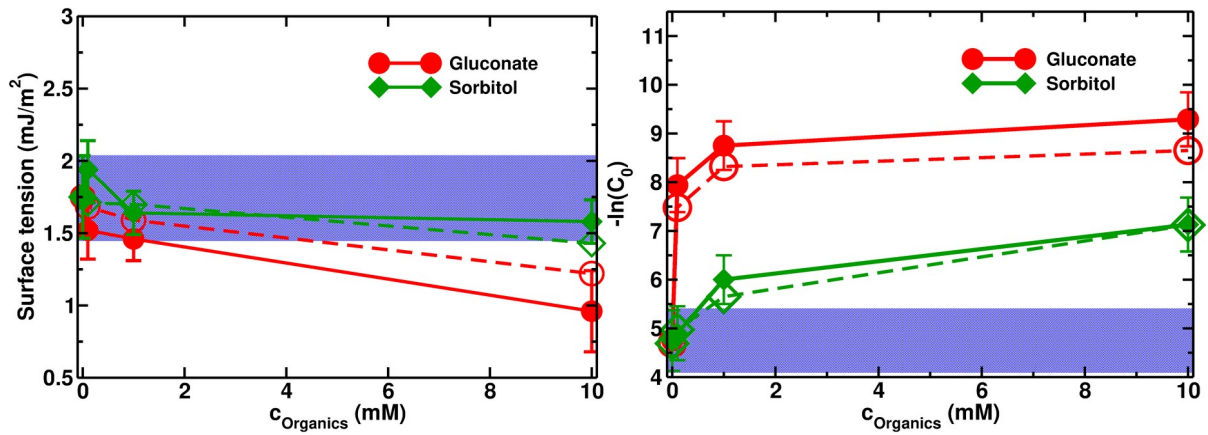

**Figure S5.** Variation of the interface tension of amorphous droplets (Left) and kinetic prefactor (Right) with the concentration of gluconate and sorbitol in the supersaturated solution. Filled symbols provide the results as obtained from Eq. 2 and empty symbols those from Eqs. 5-6, see main manuscript for more details. The error associated with the results shown by the empty symbol are not larger than the size of the symbols.

**Table S2:** Organic and elemental composition of the supersaturated solutions prospected in the precipitation experiments together with the calculated pH and saturation index (SI) with respect to β-C-S-H. Concentrations are provided in millimolar (mM); Glu: gluconate; Sorb: sorbitol; Gal: galactitol; Man: mannitol.

| pH | [Si] | [Glu] | [Sorb] | [Gal] | [Man] | [Cl] | [Na] | [Cl] | SI |
|---|---|---|---|---|---|---|---|---|---|
| 12.7958 | 0.05 | 0 | 0 | 0 | 0 | 15 | 105.1 | 55 | 1.51 |
| 12.7958 | 0.07 | 0 | 0 | 0 | 0 | 15 | 105.14 | 55 | 1.8 |
| 12.7959 | 0.1 | 0 | 0 | 0 | 0 | 15 | 105.2 | 55 | 2.11 |
| 12.796 | 0.15 | 0 | 0 | 0 | 0 | 15 | 105.3 | 55 | 2.47 |
| 12.7961 | 0.2 | 0 | 0 | 0 | 0 | 15 | 105.4 | 55 | 2.72 |
| 12.7962 | 0.25 | 0 | 0 | 0 | 0 | 15 | 105.5 | 55 | 2.91 |
| 12.7963 | 0.3 | 0 | 0 | 0 | 0 | 15 | 105.6 | 55 | 3.07 |
| 12.7964 | 0.35 | 0 | 0 | 0 | 0 | 15 | 105.7 | 55 | 3.2 |
| 12.748 | 0.05 | 10 | 0 | 0 | 0 | 15 | 115.1 | 55 | 0.53 |
| 12.7481 | 0.08 | 10 | 0 | 0 | 0 | 15 | 115.14 | 55 | 0.85 |
| 12.7483 | 0.1 | 10 | 0 | 0 | 0 | 15 | 115.2 | 55 | 1.00 |
| 12.7486 | 0.15 | 10 | 0 | 0 | 0 | 15 | 115.3 | 55 | 1.26 |
| 12.7489 | 0.2 | 10 | 0 | 0 | 0 | 15 | 115.4 | 55 | 1.43 |
| 12.7493 | 0.25 | 10 | 0 | 0 | 0 | 15 | 115.5 | 55 | 1.56 |
| 12.7497 | 0.3 | 10 | 0 | 0 | 0 | 15 | 115.6 | 55 | 1.67 |
| 12.7902 | 0.05 | 0 | 10 | 0 | 0 | 15 | 105.1 | 55 | 1.12 |
| 12.7902 | 0.07 | 0 | 10 | 0 | 0 | 15 | 105.14 | 55 | 1.35 |
| 12.7902 | 0.1 | 0 | 10 | 0 | 0 | 15 | 105.2 | 55 | 1.59 |
| 12.7903 | 0.15 | 0 | 10 | 0 | 0 | 15 | 105.3 | 55 | 1.86 |
| 12.7903 | 0.2 | 0 | 10 | 0 | 0 | 15 | 105.4 | 55 | 2.04 |
| 12.7904 | 0.25 | 0 | 10 | 0 | 0 | 15 | 105.5 | 55 | 2.17 |
| 12.7908 | 0.05 | 0 | 0 | 10 | 0 | 15 | 105.1 | 55 | 1.39 |
| 12.7909 | 0.1 | 0 | 0 | 10 | 0 | 15 | 105.2 | 55 | 1.97 |
| 12.791 | 0.15 | 0 | 0 | 10 | 0 | 15 | 105.3 | 55 | 2.31 |
| 12.7911 | 0.2 | 0 | 0 | 10 | 0 | 15 | 105.4 | 55 | 2.55 |
| 12.7912 | 0.25 | 0 | 0 | 10 | 0 | 15 | 105.5 | 55 | 2.73 |
| 12.7917 | 0.05 | 0 | 0 | 0 | 10 | 15 | 105.1 | 55 | 1.45 |
| 12.7918 | 0.1 | 0 | 0 | 0 | 10 | 15 | 105.2 | 55 | 2.05 |
| 12.7919 | 0.15 | 0 | 0 | 0 | 10 | 15 | 105.3 | 55 | 2.4 |
| 12.792 | 0.2 | 0 | 0 | 0 | 10 | 15 | 105.4 | 55 | 2.65 |
| 12.7922 | 0.25 | 0 | 0 | 0 | 10 | 15 | 105.5 | 55 | 2.84 |
| 12.7915 | 0.05 | 1 | 0 | 0 | 0 | 15 | 106.1 | 55 | 1.44 |
| 12.7916 | 0.07 | 1 | 0 | 0 | 0 | 15 | 106.14 | 55 | 1.72 |
| 12.7916 | 0.08 | 1 | 0 | 0 | 0 | 15 | 106.16 | 55 | 1.83 |
| 12.7916 | 0.1 | 1 | 0 | 0 | 0 | 15 | 106.2 | 55 | 2.02 |
| 12.7918 | 0.15 | 1 | 0 | 0 | 0 | 15 | 106.3 | 55 | 2.35 |
| 12.7919 | 0.2 | 1 | 0 | 0 | 0 | 15 | 106.4 | 55 | 2.59 |
| 12.7921 | 0.25 | 1 | 0 | 0 | 0 | 15 | 106.5 | 55 | 2.77 |
| 12.7922 | 0.3 | 1 | 0 | 0 | 0 | 15 | 106.6 | 55 | 2.91 |
| 12.7924 | 0.35 | 1 | 0 | 0 | 0 | 15 | 106.7 | 55 | 3.03 |
| 12.7955 | 0.05 | 0.1 | 0 | 0 | 0 | 15 | 105.2 | 55 | 1.51 |
| 12.7956 | 0.07 | 0.1 | 0 | 0 | 0 | 15 | 105.24 | 55 | 1.8 |
| 12.7956 | 0.1 | 0.1 | 0 | 0 | 0 | 15 | 105.3 | 55 | 2.11 |
| 12.7958 | 0.15 | 0.1 | 0 | 0 | 0 | 15 | 105.4 | 55 | 2.46 |
| 12.7959 | 0.2 | 0.1 | 0 | 0 | 0 | 15 | 105.5 | 55 | 2.71 |
| 12.796 | 0.25 | 0.1 | 0 | 0 | 0 | 15 | 105.6 | 55 | 2.9 |
| 12.7961 | 0.3 | 0.1 | 0 | 0 | 0 | 15 | 105.7 | 55 | 3.06 |
| 12.7962 | 0.35 | 0.1 | 0 | 0 | 0 | 15 | 105.8 | 55 | 3.19 |
| 12.7953 | 0.05 | 0 | 1 | 0 | 0 | 15 | 105.1 | 55 | 1.50 |
| 12.7953 | 0.07 | 0 | 1 | 0 | 0 | 15 | 105.14 | 55 | 1.79 |
| 12.7954 | 0.1 | 0 | 1 | 0 | 0 | 15 | 105.2 | 55 | 2.09 |
| 12.7955 | 0.15 | 0 | 1 | 0 | 0 | 15 | 105.3 | 55 | 2.44 |
| 12.7956 | 0.2 | 0 | 1 | 0 | 0 | 15 | 105.4 | 55 | 2.69 |
| 12.7957 | 0.25 | 0 | 1 | 0 | 0 | 15 | 105.5 | 55 | 2.88 |
| 12.7958 | 0.3 | 0 | 1 | 0 | 0 | 15 | 105.6 | 55 | 3.03 |
| 12.7959 | 0.35 | 0 | 1 | 0 | 0 | 15 | 105.7 | 55 | 3.16 |
| 12.7958 | 0.05 | 0 | 0.1 | 0 | 0 | 15 | 105.1 | 55 | 1.51 |
| 12.7958 | 0.07 | 0 | 0.1 | 0 | 0 | 15 | 105.14 | 55 | 1.8 |
| 12.7959 | 0.1 | 0 | 0.1 | 0 | 0 | 15 | 105.2 | 55 | 2.11 |
| 12.796 | 0.15 | 0 | 0.1 | 0 | 0 | 15 | 105.3 | 55 | 2.46 |
| 12.7961 | 0.2 | 0 | 0.1 | 0 | 0 | 15 | 105.4 | 55 | 2.71 |
| 12.7962 | 0.25 | 0 | 0.1 | 0 | 0 | 15 | 105.5 | 55 | 2.91 |
| 12.7963 | 0.3 | 0 | 0.1 | 0 | 0 | 15 | 105.6 | 55 | 3.07 |
| 12.7964 | 0.35 | 0 | 0.1 | 0 | 0 | 15 | 105.7 | 55 | 3.2 |

**Table S3:** Napierian logarithm of the measured induction time (in s) for the imposed organics and silicate concentrations (in mM) of the supersaturated solutions and calculated $SI^{-2}$ with respect to β-C-S-H

| [Glu] | [Sorb] | [Gal] | [Man] | [Si] | $1/SI^2$ | $\ln(t_{ind})$ |
|---|---|---|---|---|---|---|
| 10 | 0 | 0 | 0 | 0.08 | 1.37 | 10.13 |
| 10 | 0 | 0 | 0 | 0.08 | 1.37 | 10.22 |
| 10 | 0 | 0 | 0 | 0.1 | 1 | 10.6 |
| 10 | 0 | 0 | 0 | 0.1 | 1 | 10.46 |
| 10 | 0 | 0 | 0 | 0.1 | 1 | 10.82 |
| 10 | 0 | 0 | 0 | 0.15 | 0.63 | 10.09 |
| 10 | 0 | 0 | 0 | 0.15 | 0.63 | 8.5 |
| 10 | 0 | 0 | 0 | 0.2 | 0.49 | 9.77 |
| 10 | 0 | 0 | 0 | 0.2 | 0.49 | 10.13 |
| 10 | 0 | 0 | 0 | 0.2 | 0.49 | 9.9 |
| 10 | 0 | 0 | 0 | 0.25 | 0.41 | 9.55 |
| 10 | 0 | 0 | 0 | 0.25 | 0.41 | 9.47 |
| 10 | 0 | 0 | 0 | 0.3 | 0.35 | 2.48 |
| 10 | 0 | 0 | 0 | 0.3 | 0.35 | 2.63 |
| 0 | 10 | 0 | 0 | 0.05 | 0.8 | 8.61 |
| 0 | 10 | 0 | 0 | 0.05 | 0.8 | 9.43 |
| 0 | 10 | 0 | 0 | 0.05 | 0.8 | 9.62 |
| 0 | 10 | 0 | 0 | 0.07 | 0.55 | 8.41 |
| 0 | 10 | 0 | 0 | 0.07 | 0.55 | 9 |
| 0 | 10 | 0 | 0 | 0.07 | 0.55 | 9.21 |
| 0 | 10 | 0 | 0 | 0.1 | 0.39 | 8.01 |
| 0 | 10 | 0 | 0 | 0.1 | 0.39 | 8.01 |
| 0 | 10 | 0 | 0 | 0.1 | 0.39 | 8.48 |
| 0 | 10 | 0 | 0 | 0.15 | 0.29 | 7.17 |
| 0 | 10 | 0 | 0 | 0.15 | 0.29 | 7.24 |
| 0 | 10 | 0 | 0 | 0.2 | 0.24 | 3.81 |
| 0 | 10 | 0 | 0 | 0.2 | 0.24 | 4.09 |
| 0 | 10 | 0 | 0 | 0.2 | 0.24 | 3.04 |
| 0 | 10 | 0 | 0 | 0.2 | 0.24 | 3 |
| 0 | 10 | 0 | 0 | 0.25 | 0.21 | 3.18 |
| 0 | 10 | 0 | 0 | 0.25 | 0.21 | 3.2 |
| 0 | 10 | 0 | 0 | 0.25 | 0.21 | 2.29 |
| 0 | 10 | 0 | 0 | 0.25 | 0.21 | 2.3 |
| 0 | 0 | 10 | 0 | 0.05 | 0.52 | 10.62 |
| 0 | 0 | 10 | 0 | 0.05 | 0.52 | 9.21 |
| 0 | 0 | 10 | 0 | 0.05 | 0.52 | 8.7 |
| 0 | 0 | 10 | 0 | 0.05 | 0.52 | 8.29 |
| 0 | 0 | 10 | 0 | 0.1 | 0.26 | 8.27 |
| 0 | 0 | 10 | 0 | 0.1 | 0.26 | 7.52 |
| 0 | 0 | 10 | 0 | 0.1 | 0.26 | 7.6 |
| 0 | 0 | 10 | 0 | 0.15 | 0.19 | 7.35 |
| 0 | 0 | 10 | 0 | 0.15 | 0.19 | 7.5 |
| 0 | 0 | 10 | 0 | 0.2 | 0.15 | 4.09 |
| 0 | 0 | 10 | 0 | 0.2 | 0.15 | 4.09 |
| 0 | 0 | 10 | 0 | 0.25 | 0.13 | 1.7 |
| 0 | 0 | 10 | 0 | 0.25 | 0.13 | 1.65 |
| 0 | 0 | 0 | 10 | 0.05 | 0.48 | 8.61 |
| 0 | 0 | 0 | 10 | 0.05 | 0.48 | 10.54 |
| 0 | 0 | 0 | 10 | 0.05 | 0.48 | 10.16 |
| 0 | 0 | 0 | 10 | 0.1 | 0.24 | 7.63 |
| 0 | 0 | 0 | 10 | 0.1 | 0.24 | 7.71 |
| 0 | 0 | 0 | 10 | 0.1 | 0.24 | 7.74 |
| 0 | 0 | 0 | 10 | 0.15 | 0.17 | 7.07 |
| 0 | 0 | 0 | 10 | 0.15 | 0.17 | 7.09 |
| 0 | 0 | 0 | 10 | 0.15 | 0.17 | 7.05 |
| 0 | 0 | 0 | 10 | 0.2 | 0.14 | 3 |
| 0 | 0 | 0 | 10 | 0.2 | 0.14 | 3.01 |
| 0 | 0 | 0 | 10 | 0.2 | 0.14 | 3.02 |
| 0 | 0 | 0 | 10 | 0.25 | 0.12 | 2.3 |
| 0 | 0 | 0 | 10 | 0.25 | 0.12 | 2.2 |
| 1 | 0 | 0 | 0 | 0.05 | 0.34 | 10.6 |
| 1 | 0 | 0 | 0 | 0.07 | 0.3 | 8.01 |
| 1 | 0 | 0 | 0 | 0.07 | 0.3 | 9.43 |
| 1 | 0 | 0 | 0 | 0.07 | 0.3 | 11.29 |
| 1 | 0 | 0 | 0 | 0.07 | 0.3 | 10.6 |
| 1 | 0 | 0 | 0 | 0.08 | 0.25 | 9.62 |
| 1 | 0 | 0 | 0 | 0.08 | 0.25 | 8.92 |
| 1 | 0 | 0 | 0 | 0.1 | 0.18 | 9.9 |
| 1 | 0 | 0 | 0 | 0.1 | 0.18 | 8.52 |
| 1 | 0 | 0 | 0 | 0.1 | 0.18 | 9.43 |
| 1 | 0 | 0 | 0 | 0.1 | 0.18 | 8.92 |
| 1 | 0 | 0 | 0 | 0.15 | 0.15 | 7.31 |
| 1 | 0 | 0 | 0 | 0.15 | 0.15 | 7.6 |
| 1 | 0 | 0 | 0 | 0.2 | 0.13 | 4.61 |
| 1 | 0 | 0 | 0 | 0.2 | 0.13 | 4.09 |
| 1 | 0 | 0 | 0 | 0.25 | 0.12 | 2.7 |
| 1 | 0 | 0 | 0 | 0.25 | 0.12 | 2.63 |
| 1 | 0 | 0 | 0 | 0.3 | 0.11 | 3 |
| 1 | 0 | 0 | 0 | 0.3 | 0.11 | 3 |
| 1 | 0 | 0 | 0 | 0.35 | 0.1 | 3.91 |
| 1 | 0 | 0 | 0 | 0.35 | 0.1 | 3.4 |
| 0.1 | 0 | 0 | 0 | 0.05 | 0.44 | 10.13 |
| 0.1 | 0 | 0 | 0 | 0.05 | 0.44 | 9.39 |
| 0.1 | 0 | 0 | 0 | 0.05 | 0.44 | 8.7 |
| 0.1 | 0 | 0 | 0 | 0.05 | 0.44 | 9.9 |
| 0.1 | 0 | 0 | 0 | 0.05 | 0.44 | 8 |
| 0.1 | 0 | 0 | 0 | 0.05 | 0.44 | 8.99 |
| 0.1 | 0 | 0 | 0 | 0.1 | 0.23 | 8.96 |
| 0.1 | 0 | 0 | 0 | 0.1 | 0.23 | 8.96 |
| 0.1 | 0 | 0 | 0 | 0.1 | 0.23 | 9.04 |
| 0.1 | 0 | 0 | 0 | 0.15 | 0.17 | 8.78 |
| 0.1 | 0 | 0 | 0 | 0.15 | 0.17 | 8.62 |
| 0.1 | 0 | 0 | 0 | 0.15 | 0.17 | 8.29 |
| 0.1 | 0 | 0 | 0 | 0.2 | 0.14 | 8.06 |
| 0.1 | 0 | 0 | 0 | 0.2 | 0.14 | 8.07 |
| 0.1 | 0 | 0 | 0 | 0.25 | 0.12 | 3.6 |
| 0.1 | 0 | 0 | 0 | 0.25 | 0.12 | 3.26 |
| 0.1 | 0 | 0 | 0 | 0.35 | 0.1 | 3.69 |
| 0.1 | 0 | 0 | 0 | 0.35 | 0.1 | 3.76 |
| 0.1 | 0 | 0 | 0 | 0.4 | 0.09 | 3.4 |
| 0.1 | 0 | 0 | 0 | 0.4 | 0.09 | 3.5 |
| 0.1 | 0 | 0 | 0 | 0.45 | 0.62 | 9.79 |
| 0.1 | 0 | 0 | 0 | 0.45 | 0.62 | 9.39 |
| 0.1 | 0 | 0 | 0 | 0.45 | 0.62 | 8.29 |
| 0 | 1 | 0 | 0 | 0.1 | 0.23 | 6.9 |
| 0 | 1 | 0 | 0 | 0.1 | 0.23 | 6.9 |
| 0 | 1 | 0 | 0 | 0.1 | 0.23 | 7.31 |
| 0 | 1 | 0 | 0 | 0.15 | 0.17 | 6.91 |
| 0 | 1 | 0 | 0 | 0.15 | 0.17 | 6.62 |
| 0 | 1 | 0 | 0 | 0.15 | 0.17 | 6.62 |
| 0 | 1 | 0 | 0 | 0.2 | 0.14 | 5.01 |
| 0 | 1 | 0 | 0 | 0.2 | 0.14 | 5.01 |
| 0 | 1 | 0 | 0 | 0.2 | 0.14 | 5.99 |
| 0 | 1 | 0 | 0 | 0.25 | 0.12 | 4.09 |
| 0 | 1 | 0 | 0 | 0.25 | 0.12 | 4.17 |
| 0 | 1 | 0 | 0 | 0.3 | 0.11 | 3.46 |
| 0 | 1 | 0 | 0 | 0.3 | 0.11 | 3.57 |
| 0 | 1 | 0 | 0 | 0.35 | 0.1 | 1.38 |
| 0 | 1 | 0 | 0 | 0.35 | 0.1 | 1.09 |
| 0 | 0.1 | 0 | 0 | 0.05 | 0.44 | 9.29 |
| 0 | 0.1 | 0 | 0 | 0.05 | 0.44 | 9.16 |
| 0 | 0.1 | 0 | 0 | 0.05 | 0.44 | 9.33 |
| 0 | 0.1 | 0 | 0 | 0.1 | 0.22 | 6.68 |
| 0 | 0.1 | 0 | 0 | 0.1 | 0.22 | 6.68 |
| 0 | 0.1 | 0 | 0 | 0.1 | 0.22 | 7.5 |
| 0 | 0.1 | 0 | 0 | 0.15 | 0.17 | 5.86 |
| 0 | 0.1 | 0 | 0 | 0.15 | 0.17 | 4.7 |
| 0 | 0.1 | 0 | 0 | 0.15 | 0.17 | 6.8 |
| 0 | 0.1 | 0 | 0 | 0.2 | 0.14 | 5.3 |
| 0 | 0.1 | 0 | 0 | 0.2 | 0.14 | 5.3 |
| 0 | 0.1 | 0 | 0 | 0.2 | 0.14 | 5.01 |
| 0 | 0.1 | 0 | 0 | 0.35 | 0.1 | 3.81 |
| 0 | 0.1 | 0 | 0 | 0.35 | 0.1 | 3.71 |
| 0 | 0 | 0 | 0 | 0.05 | 0.44 | 7.35 |
| 0 | 0 | 0 | 0 | 0.05 | 0.44 | 6.41 |
| 0 | 0 | 0 | 0 | 0.05 | 0.44 | 7.91 |
| 0 | 0 | 0 | 0 | 0.05 | 0.44 | 6.45 |
| 0 | 0 | 0 | 0 | 0.07 | 0.31 | 6.53 |
| 0 | 0 | 0 | 0 | 0.07 | 0.31 | 6.23 |
| 0 | 0 | 0 | 0 | 0.07 | 0.31 | 6.45 |
| 0 | 0 | 0 | 0 | 0.1 | 0.22 | 5.89 |
| 0 | 0 | 0 | 0 | 0.1 | 0.22 | 6.38 |
| 0 | 0 | 0 | 0 | 0.1 | 0.22 | 5.48 |
| 0 | 0 | 0 | 0 | 0.1 | 0.22 | 7 |
| 0 | 0 | 0 | 0 | 0.15 | 0.16 | 6.03 |
| 0 | 0 | 0 | 0 | 0.15 | 0.16 | 4.91 |
| 0 | 0 | 0 | 0 | 0.15 | 0.16 | 6.21 |
| 0 | 0 | 0 | 0 | 0.2 | 0.14 | 4.79 |
| 0 | 0 | 0 | 0 | 0.2 | 0.14 | 4.61 |
| 0 | 0 | 0 | 0 | 0.2 | 0.14 | 6.21 |
| 0 | 0 | 0 | 0 | 0.25 | 0.12 | 2.3 |
| 0 | 0 | 0 | 0 | 0.25 | 0.12 | 2.2 |